\documentclass[]{aastex63}
\usepackage{graphicx}
\usepackage{amsmath}
\usepackage{booktabs}
\usepackage{multirow}
\usepackage{float}
\usepackage{lineno}
\usepackage{subfigure}
\usepackage{subfloat}
\usepackage{overpic}
\usepackage{ulem}
\usepackage{longtable}
\usepackage[flushleft]{threeparttable}
\usepackage{hyperref}
\usepackage[hyphenbreaks]{breakurl}
\usepackage{CJKutf8}
\usepackage[normalem]{ulem}

\shorttitle{Advancing Identification of GRBs}
\shortauthors{Zhang et al.}

\begin{document}
\begin{CJK*}{UTF8}{gbsn}

\title{Advancing Identification method of Gamma-Ray Bursts with Data and Feature Enhancement}


\def\Tongji{College of Electronic and Information Engineering, Tongji University, Shanghai 201804, China}

\def\HighEnergy{Key Laboratory of Particle Astrophysics, Chinese Academy of Sciences, Beijing 100049, China, \url{libing@ihep.ac.cn, rzgui@tongji.edu.cn, xiongsl@ihep.ac.cn}}

\def\Tsinghua{Department of Astronomy, Tsinghua University, Beijing 100048, China,
\url{libing307@tsinghua.edu.cn}}

\def\Guokeda{University of Chinese Academy of Sciences, Beijing 100049, Beijing, China}

\def\jinggangshan{Department of Physics, Jinggangshan University, Jiangxi Province, Ji'an 343009, China}

\def\Nanjing{School of Astronomy and Space Science, Nanjing University, Nanjing 210023, China}

\def\Guangxi{Guangxi Key Laboratory for Relativistic Astrophysics, Nanning 530004, China}

\def\XinanJiaoTong{School of Computing and Artificial Intelligence, Southwest
Jiaotong University, Chengdu 611756, China}

\def\HeBeiNormal{College of Physics and Hebei Key Laboratory of Photophysics Research and Application, Hebei Normal University, Shijiazhuang, Hebei 050024, China}

\def\Dezhou{School of Computer and Information, Dezhou University, Dezhou 253023, China}

\def\BeijingNormal{Department of Astronomy, Beijing Normal University, Beijing 100875, China}

\author[0000-0002-8097-3616]{Peng Zhang (张鹏)}
\affiliation{\Tongji}
\affiliation{\HighEnergy{}}

\author[0000-0002-0238-834X]{Bing Li (李兵)}
\affiliation{\HighEnergy{}}
\affiliation{\Tsinghua{}}
\affiliation{\Guangxi{}}

\author{Ren-Zhou Gui (桂任舟)}
\affiliation{\Tongji}

\author[0000-0002-4771-7653]{Shao-Lin Xiong (熊少林)}
\affiliation{\HighEnergy{}}

\author[0000-0001-7959-3387]{Yu Wang (王瑜)}
\affiliation{ICRA, Dip. di Fisica, Sapienza Universit\`a di Roma, Piazzale Aldo Moro 5, I-00185 Roma, Italy}
\affiliation{ICRANet, Piazza della Repubblica 10, 65122 Pescara, Italy}
\affiliation{INAF -- Osservatorio Astronomico d'Abruzzo, Via M. Maggini snc, I-64100, Teramo, Italy}

\author{Shi-Jie Zheng (郑世界)}
\affiliation{\HighEnergy{}}

\author{Guang-Cheng Xiao (肖广成)}
\affiliation{\jinggangshan}

\author{Xiao-Bo Li (李小波)}
\affiliation{\HighEnergy{}}

\author{Yue Huang (黄跃)}
\affiliation{\HighEnergy{}}

\author[0009-0008-8053-2985]{Chen-Wei Wang (王晨巍)}
\affiliation{\HighEnergy{}}

\author[0009-0004-1887-4686]{Jia-Cong Liu (刘佳聪)}
\affiliation{\HighEnergy{}}

\author[0000-0001-5348-7033]{Yan-Qiu Zhang (张艳秋)}
\affiliation{\HighEnergy{}}

\author[0000-0001-8664-5085]{Wang-Chen Xue (薛王陈)}
\affiliation{\HighEnergy{}}

\author[0009-0001-7226-2355]{Chao Zheng (郑超)}
\affiliation{\HighEnergy{}}

\author[0009-0008-5068-3504]{Yue Wang (王悦)}
\affiliation{\HighEnergy{}}

\begin{abstract}
Gamma-ray bursts (GRBs) are challenging to identify due to their transient 
nature, complex temporal profiles, and limited observational datasets. 
We address this with a one-dimensional convolutional neural network 
integrated with an Adaptive Frequency Feature Enhancement module and 
physics-informed data augmentation.
Our framework generates 100,000 synthetic GRB samples, expanding training 
data diversity and volume while preserving physical fidelity-especially 
for low-significance events. 
The model achieves 97.46\% classification accuracy, outperforming all 
tested variants with conventional enhancement modules, highlighting
enhanced domain-specific feature capture. 
Feature visualization shows model focuses on deep-seated morphological features 
and confirms the capability of extracting physically meaningful burst characteristics. 
Dimensionality reduction and clustering reveal GRBs with similar morphologies 
or progenitor origins cluster in the feature space, linking learned features 
to physical properties. This perhaps offers a novel diagnostic tool for identifying 
kilonova- and supernova-associated GRB candidates, establishing criteria 
to enhance multi-messenger early-warning systems.
The framework aids current time-domain surveys, generalizes to other rare 
transients, and advances automated detection in large-volume observational data. 
\end{abstract}


\section{Introduction}
\label{sect:intro}

Gamma-ray bursts are among the most energetic phenomena in the universe, 
releasing enormous amounts of energy in the form of gamma rays over remarkably 
short time frames. Despite decades of remarkable progress in observations and 
theoretical investigations, their origins remain elusive
~\citep{Zhang2011CRP, KumarZhang2015PhR, Meszaros2019MmSAI, Peer2024Galax}, 
making GRB identification and classification a critical focus of contemporary 
astrophysical research~\citep{lvhoujun2010ApJ, ZhangShuai2022ApJ, Sun2023Arxiv, Wang2024Arxiv}.
The rise of multi-messenger and multi-band astronomy further underscores the 
need for efficient GRB detection \citep{Margutti2021ARAA, Rudolph2023ApJ}: 
timely and accurate detection enables follow-up observations of afterglows, 
host galaxy localization, characterization of associated counterparts, 
and constraints on GRB intrinsic properties.

GRB light curves often exhibit irregular, multi-peaked structures 
that encode rich physical information about their outbursts. 
While traditional classification criteria (e.g., duration, hardness)
have been widely adopted, emerging evidence highlights the need for additional 
discriminants to address the "iceberg effect"—a phenomenon where low-signal-to-noise 
ratio (low-SNR) GRBs are partially submerged in background noise, disproportionately 
impacting short-duration events \citep{2014MNRAS.442.1922L, iceberg-effects_2022Moss}.
Traditional burst search algorithms identify GRBs by detecting signals in 
multi-time-bin, multi-energy-band light curves where the SNR exceeds a predefined 
threshold above the 
background \citep{fermi_blind_search, fermi_target_search, search_grb_hxmt_cai}. 
However, these methods face significant challenges, including accurate background 
estimation and optimal threshold selection, limiting their effectiveness in 
identifying faint or complex GRB events. 
These limitations motivate the development of advanced techniques to enhance 
GRB detection and classification accuracy.

In recent years, machine learning (ML) has emerged as an experimental tool for GRB 
detection. For example, \cite{relate_search_GRB_DTW} developed an ML algorithm for 
automated detection of GRB-like events using AstroSat/CZTI data, combining 
density-based spatial clustering with dynamic time warping to demonstrate ML’s 
versatility and robustness in GRB identification. 
Among ML architectures, convolutional neural networks (CNNs)—a specialized deep 
learning (DL) framework—excel in pattern recognition 
tasks \citep{Alzubaidi2021ReviewOD, Taye2023TheoreticalUO}, and have driven a 
paradigm shift in astrophysical data analysis by autonomously extracting features 
from multi-dimensional, multi-modal datasets.
In GRB research, CNN-based approaches have achieved notable successes.
\cite{DL_detect_GRB_intensity_map} developed a real-time CNN pipeline for 
AGILE-GRID intensity maps, detecting 21 GRBs at 3\(\sigma\) significance compared to 
only two detections via conventional SNR-based searches.
\cite{DL_autodencoder_detect_grb} used a CNN autoencoder to reconstruct background 
light curves and identify 72 previously uncataloged AGILE GRBs.
Additionally, \citep{ParmiggianiarX240402107P} and \citet{nn_search_long_faint_burst} 
leveraged neural networks to predict background count rates and identify faint, 
long-duration GRBs.

Notably, GRB identification relies on extracting meaningful features from light curves 
with complex temporal and spectral characteristics. 
While traditional 1D-CNNs effectively capture local temporal patterns, 
their inability to model spectral-temporal correlations limits their utility in GRB 
analysis—critical GRB features are jointly encoded in both 
domains \citep{GRB-fft_2016Guidorzi, GRB-fft_2021Tarnopolski, GRB-fft_2024Zhou}. 
To address this, recent time series classification advances have introduced architectures 
that explicitly model frequency information, including Wavelet-based decomposition 
(e.g., T-WaveNet \citep{CNN_freq_liu_2020}), 2D temporal representations 
(e.g., TimesNet \citep{CNN_freq_wu_2022}), hybrid Fourier-wavelet transformers 
(e.g., FEDformer \citep{CNN_freq_zhou_2022}, WFTNet \citep{CNN_freq_liu_2024}), 
and adaptive spectral blocks (e.g., TSLANet \citep{model_block_ASB}). 
These state-of-the-art approaches confirm that spectral-temporal modeling enhances 
feature extraction, underscoring the need for GRB identification methods that 
integrate both temporal dynamics and frequency-domain characteristics.

Complementary to classification, dimensionality reduction techniques have provided 
valuable insights into GRB diversity. 
For instance, \citet{DownDimGRB_jesperen2020} used t-distributed stochastic neighbor 
embedding (t-SNE) to categorize Swift/BAT GRBs to separate supernova-associated and 
kilonova-associated GRBs into long and short categories.
\cite{ML-downdim-GRB-EE_2023Garcia} identified 7 new extended emission (EE) GRB 
candidates via dimensionality reduction. 
\cite{2024MNRAS.532.1434Z} and \citet{DownDimGRB_Negro2025} used Uniform Manifold 
Approximation and Projection (UMAP) or t-SNE to reveal GRB clusters that transcend 
traditional duration-based classification, including distinct subgroups of 
kilonova-related GRBs. 
Similarly, \cite{2024MNRAS.527.4272C_Chen} and \citet{DownDimGRB_Chen2025} 
demonstrated robust bimodal GRB clustering and refuted intermediate GRB classes 
using unsupervised ML and dimensionality reduction. 
Collectively, these studies highlight the utility of such techniques for 
analyzing burst similarities and refining classification schemes.

Despite these advances, DL models face critical limitations in GRB 
research. 
First, data scarcity persists: the largest available GRB dataset (Fermi/GBM, ~3900 events) 
remains insufficient for robust DL training \citep{DL_identify_grb_by_peng}, 
due to limited labeled observations, skewed distribution (underrepresentation of rare, 
short-duration, and low-SNR GRBs), and narrow coverage of physical properties 
(e.g., redshift, energy bands, progenitors). 
While data augmentation has been explored to mitigate 
this \citep{DL_identify_grb_by_peng, DataAug_2019BigDataSurvey}, 
existing datasets still lack the feature richness needed to avoid overfitting. 
Second, uncertainty quantification (UQ) is underdeveloped: model uncertainty stems 
from limited training data, the "black-box" nature of CNNs, and observational 
noise \citep{Nemani2023MSSP20510796N}, yet UQ is critical for distinguishing 
high-confidence detections from ambiguous candidates—misclassification of which 
could lead to missed follow-up observations or erroneous inferences about 
progenitor origins.

To address these limitations, we present an integrated framework that combines 
physics-informed data augmentation (to enhance sample diversity) with a novel 
frequency-adaptive feature enhancement module (to capture spectral-temporal 
correlations). 
In this Letter, Section \ref{sec:dataset} details data preprocessing, augmentation, 
and dataset construction. Section \ref{sec:methods} describes the model architecture, 
feature enhancement modules, training process, and feature analysis framework.
Section \ref{sec:result} presents performance metrics and classification results, 
and Section \ref{sec:discussion} discusses the implications of our approach and 
summarizes conclusions.

\section{Data augmentation and Dataset}
\label{sec:dataset}

The Fermi/GBM trigger system has been operational since 12 July 2008 
(first detected GRB: GRB 080714B). 
For this study, we utilized Fermi/GBM data spanning from 14 July 2008 to 30 June 2024, 
including 3,905 original GRBs detected by NaI detectors. 
These events are manually verified and cataloged in the Fermi/GBM Burst 
Catalog\footnote{\url{https://heasarc.gsfc.nasa.gov/W3Browse/fermi/fermigbrst.html}}.
Adopting the GRB/non-GRB extraction protocol described 
in \citet{DL_identify_grb_by_peng}, we updated the dataset by segmenting samples 
into 120-second time windows. 
To ensure complete coverage of the $T_{90}$ interval (with background included on both 
sides), 217 GRBs failing this criterion were excluded. 
For the remaining events, samples were generated based on the number of triggering 
NaI detectors, resulting in 6,189 primary GRB samples. 
Non-GRB samples—encompassing modulated background, electronic noise, and potential 
all-sky source signals—were systematically extracted from detectors’ quiescent 
(non-trigger) intervals, totaling 108,000 samples. 
This balance between classes enhances model generalization to non-burst events.

Light curves for all samples were extracted at a 64\,ms temporal resolution across 
128 energy channels and rebinned into 9 standard energy bands (25–50, 50–100, 50–300, 
100–300, 100–500, 100–900, 300–500, 300–900, 500–900 keV). 
This segmentation aligns with typical GRB photon energy distributions and Fermi/GBM 
trigger criteria \citep{fermi_gbm_fouth_catalog, search_grb_hxmt_cai}, approximating 
the photon deposition ranges of NaI detectors (without energy response correction). 
Critically, it preserves energy-dependent features such as burst morphology evolution, 
light curve shape variations, and spectral lags—enriching the diversity of 
discriminative features for model training.
To optimize input for machine learning, we applied per-band 
standardization to light curves (scaling features to zero mean and unit variance). 
This approach preserves spectral information via inter-band flux ratios, 
empirically outperforming normalization (scaling to [0,1]). 
While both methods unify feature scales to facilitate deep learning convergence, 
standardization is preferable for GRB characterization as it retains spectral 
integrity—essential for distinguishing GRBs from non-GRBs (e.g., via hardness 
evolution or spectral lags).

\begin{figure}
\centering
\includegraphics[width=0.75\textwidth, angle=0]{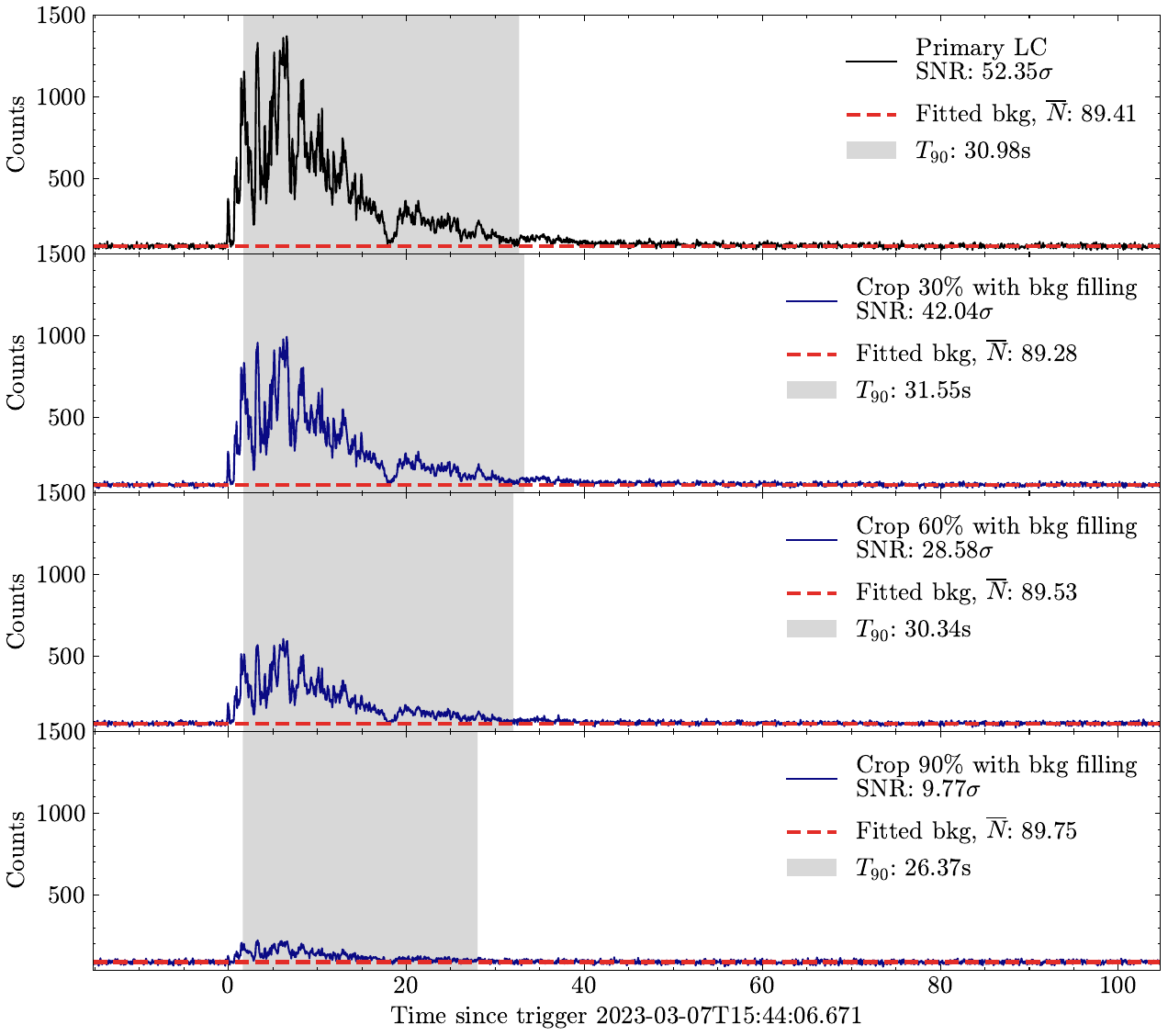}
\caption{Data augmentation procedure for GRB light curves, using GRB 230307A as observed 
by Fermi/GBM detector~n8 (full energy channel). 
The top panel presents the original light curve with a peak-SNR of 52.35\,$\sigma$, 
where the red dashed line indicates the polynomial-fitted background level 
(mean count rate $\sim$90\,counts/bin) and the shaded area marks the $T_{90}$ interval. 
The lower panels show the resulting light curves after applying count rate 
reductions of 30\%, 60\%, and 90\% to each time bin, with subsequent restoration 
of Poisson-distributed background noise at the corresponding levels.
The resulting peak SNR values are indicated for each modified light curve.}
\label{fig:crop_example}
\end{figure}

\textbf{Data augmentation:}
We first compute the full-energy band peak signal-to-noise ratio (peak-SNR) of 
primary GRB samples and fit its distribution using a log-normal function. 
This fitted distribution is then used for random sampling to generate more 
synthetic GRB samples.
Specifically, we randomly reduce the count rate of original GRB light curves 
multiple times. 
For each GRB, a single reduction factor is applied to all its 9 energy band 
light curves in each reduction, generating distinct light curves with unique 
SNRs—effectively enhancing training set diversity. 
For instance, as illustrated in Figure \ref{fig:crop_example}, we generated 
three synthetic light curves (distinct SNRs) for GRB 230307A.
This approach is analogous to resizing or contrast enhancement in affine data 
augmentation.

\begin{figure}
\centering
\includegraphics[width=0.9\textwidth, angle=0]{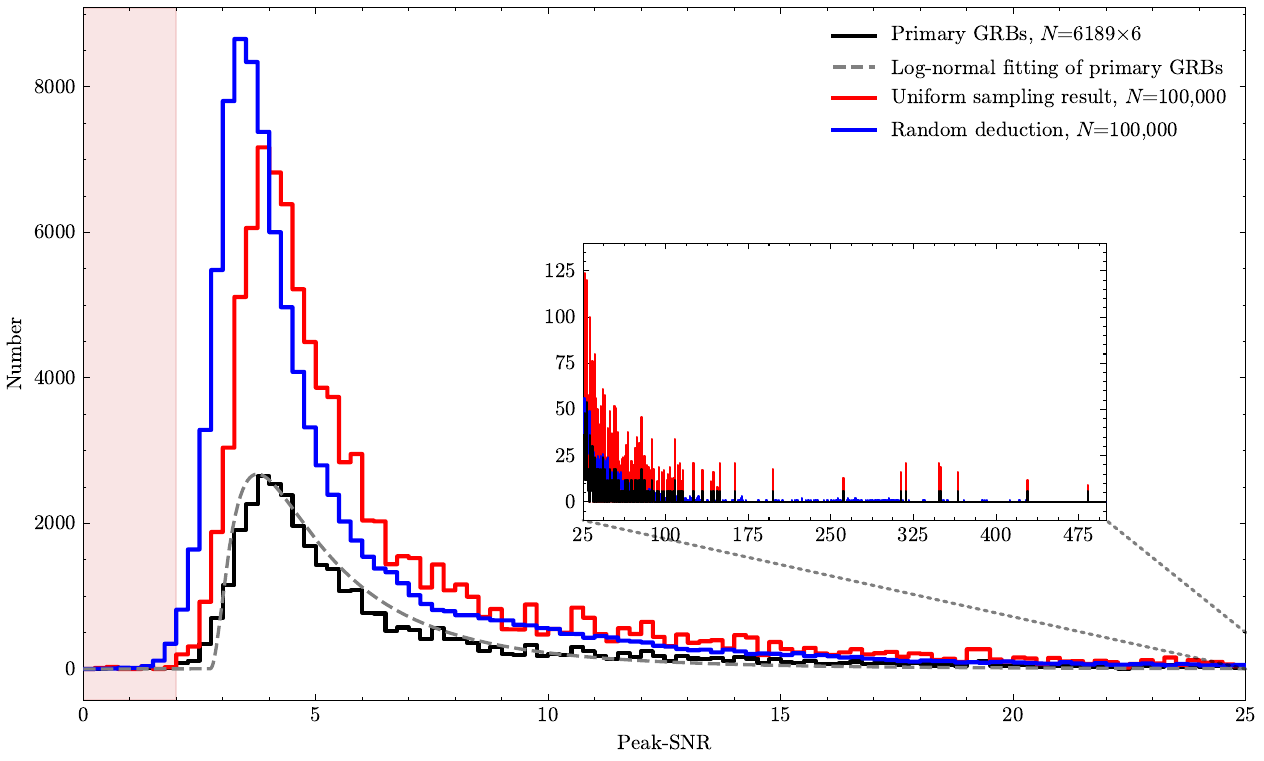}
\caption{Histogram of full-energy band peak-SNRs for GRB samples during the data 
augmentation process.
The solid black line represents the SNR distribution of the primary sample, 
scaled by a factor of 6 for visualization. 
The dashed gray line denotes the fitted log-normal distribution over the range 
[0, 25]\,$\sigma$. 
The solid red line shows the peak-SNR distribution of 100,000 uniformly sampled 
instances from the primary GRB samples.
The solid blue line represents the peak-SNR distribution of 100,000 randomly 
selected GRB samples after data augmentation. 
The pink shaded area highlights peak-SNR values below two\,$\sigma$.}
\label{fig:trainset_snr_hist}
\end{figure}

Given an input light curve $\mathbf{LC} \in \mathbb{R}^{T}$ ($\mathbb{R}$: 
real numbers; $T$: temporal length) and a background light curve
$\mathbf{LC}_{\text{bg}} \in \mathbb{R}^{T}$, the augmented light curve 
$\mathbf{LC}_{\text{new}}$ is computed as:

\begin{equation}
\mathbf{LC}_{\text{new}} = \mathbf{LC} \cdot (1 - \alpha) + \text{Poisson}(\mathbf{LC}_{\text{bg}}
\cdot \alpha),
\end{equation}

where $\alpha \in [0, 1]$ is the \textit{crop factor} controlling burst photon 
deduction proportion. 
The $\text{Poisson}(\cdot)$ function introduces stochastic noise, restoring 
matching Poisson-distributed background noise to simulate real astronomical 
random noise. 
This formula reduces burst signal significance while preserving original 
background levels, ensuring synthetic samples match real scenarios 
(e.g., orbit modulation, instrument noise).
Notably, this method reproduces different degrees of the "iceberg effect" 
in synthetic GRBs—especially effective for enriching short-duration, rare faint, 
and low-SNR GRB samples. 
Crucially, augmented samples faithfully retain inherent fine-scale temporal/
spectral substructures of original GRBs, ensuring no excessive loss of key 
physical features for GRB identification/classification.
The data augmentation process is detailed below, with statistical results 
shown in Figure \ref{fig:trainset_snr_hist}:

\begin{enumerate}
    \item Analyze full-energy band peak-SNR of 6,189 primary GRB samples 
    and fit this distribution with a log-normal function over [0,25]\,$\sigma$.

    \item Randomly sample 500,000 values from the fitted distribution and 
    divide by 25 to scale [0,25]\,$\sigma$ to [0,1], 
    yielding \textit{crop factors} ($\alpha$).
    
    \item Uniformly sample 500,000 instances from primary GRB samples and 
    pair each with a random $\alpha$, then generate 500,000 synthetic samples 
    via Equ. 1. Use the same $\alpha$ for all 9 energy bands of a GRB to 
    preserve spectral consistency.
        
    \item Remove 4,648 synthetic samples with peak-SNR below two\,$\sigma$, 
    as excessive photon subtraction makes these resemble background samples.
    
    \item Randomly draw 100,000 samples from the remainder to balance 
    GRB/non-GRB ratios in the training set\footnote{For this supervised task, 
    final sample size was determined via prior experience with similar models, 
    relying on empirical heuristics.}.
    
    \item Apply the same augmentation to non-GRB samples to align background 
    noise distributions, ensuring the model distinguishes GRBs from background 
    (not artificial Poisson noise differences).
  
\end{enumerate}

\textbf{Dataset:}
We partition preprocessed samples into training, validation, and test sets 
chronologically to ensure temporal generalization. 
Dataset details are shown in Table \ref{table:dataset}.
This chronological split alleviates temporal overfitting while preserving 
the data’s inherent flux distribution.
We maintain $\sim$ 1:1 positive-to-negative sample ratios across all splits. 
Crucially, only the training set undergoes data augmentation, while the validation/test 
sets retain original observations (primary GRB samples and non-GRB samples) 
for unbiased performance evaluation.

\begin{deluxetable*}{cccccc}[htbp]
\label{table:dataset}
\tablecaption{Description of the dataset.}
\tablehead{
\multirow{2}{*}{Dataset} & \multicolumn{2}{c}{Nu. of GRB Events} & \multicolumn{2}{c}{Nu. of Samples} & \multirow{2}{*}{Period Definition (UTC)} \\
\cmidrule[0.5pt]{2-5} 
               &Original GRBs&Primary GRBs&Synthetic GRBs&  Non-GRBs }
\startdata
Training set   & 1,899       & 6189       & 100,000      & 100,000  & 07/14/2008 - 31/12/2016 \\ 
Validation set & 842         & 2,774       & 2,774        & 4,000    & 01/01/2017 - 12/31/2019 \\
Test set       & 947         & 3,143       & 3,143        & 4,000    & 01/01/2020 - 06/31/2024 \\
\enddata
\end{deluxetable*}

\section{Model architecture, training, and extension}
\label{sec:methods}

\subsection{Architecture of Neural Networks}
\label{sec:nework_architectures} 
Each sample in our dataset consists of light curves spanning 9 energy bands, 
which are naturally suited to be treated as time series data. 
This makes a 1D CNN the appropriate choice for feature extraction and classification. 
Our model adopts ResNet \citep{model_resnet} as its backbone, which relies on 
a sequential stack of four convolutional units (Conv Units) to progressively 
extract hierarchical features from the input light curves. 
A key advantage of ResNet is its integration of residual connections, 
which effectively alleviate the gradient vanishing problem that commonly arises in 
deep neural networks and enable stable training of deeper architectures.
To address the limitation that CNNs lack the ability to distinguish subtle 
frequency variations, we propose an Adaptive Frequency Feature Enhancement module, 
AFFE, which is integrated into the last part of each Conv Unit. 
This module explicitly enhances frequency-domain features to improve the model's 
GRB identification capability, and it also adaptively weights and filters frequency 
components, enabling the model to focus on the most discriminative features for 
GRB recognition. 
This design is particularly critical for distinguishing GRBs from background noise, 
as frequency-domain analysis can uncover patterns that remain obscured in the time domain.

\begin{figure}[h]
	\begin{minipage}{\linewidth}
		\vspace{0pt}
		\centerline{\includegraphics[width=\textwidth]{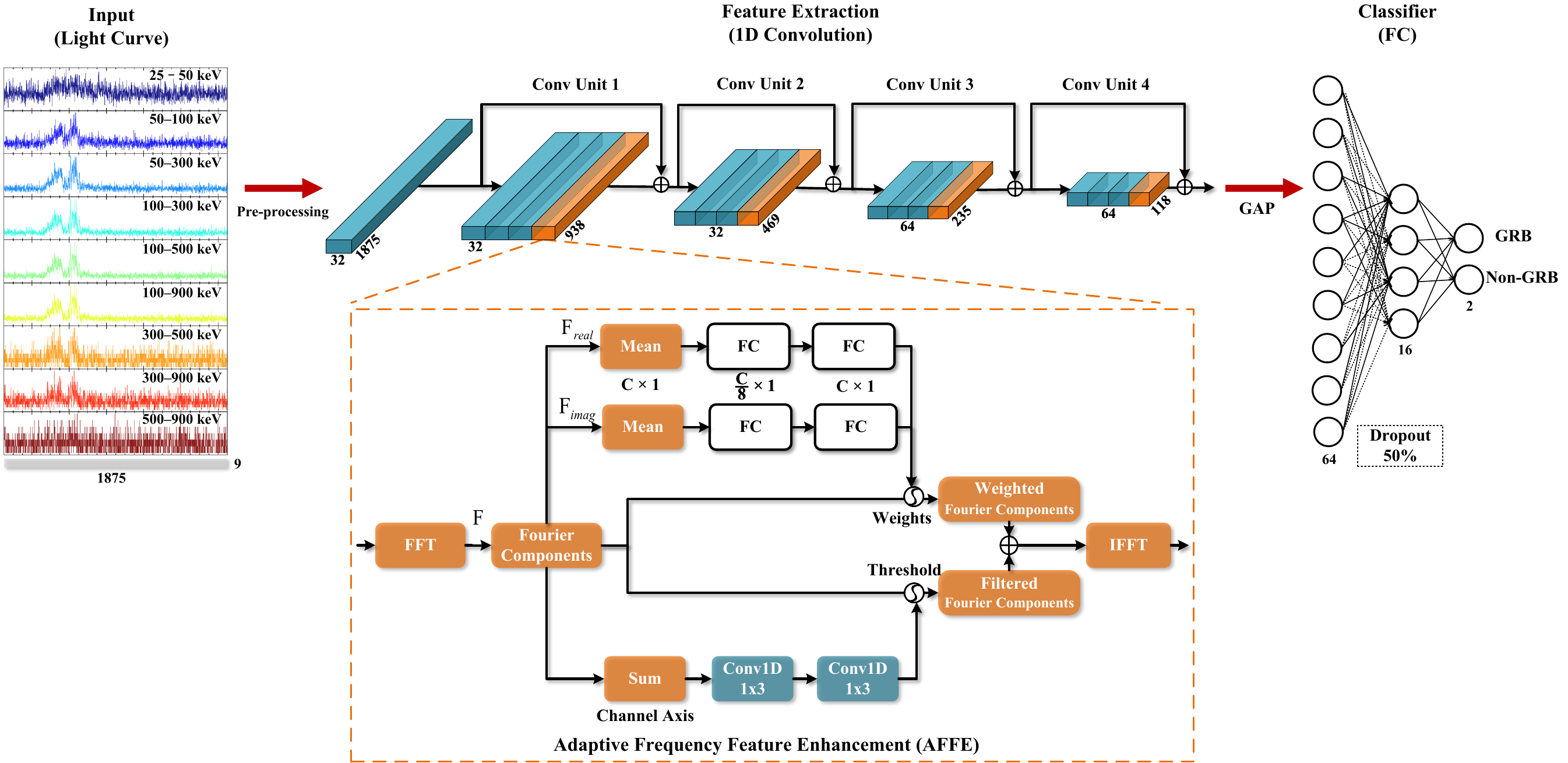}}
	\end{minipage}
	\caption{Schematic diagram of the ResNet+AFFE model architecture and the feature 
    transformation process across layers. 
    The input light curves are colored differently to distinguish the nine energy bands
    (This color distinction is solely for intuitively 
    distinguishing each energy band to improve the clarity of the diagram, 
    with no additional physical implications.). 
    The numbers inside each Conv unit indicate changes in feature map size and channel 
    dimension. The yellow modules represent the AFFE modules, and the dashed box at the 
    bottom details an implementation of the AFFE module.}
\label{fig:network_architectures}
\end{figure}

After the convolutional layers (including Conv Units with AFFE modules) complete 
feature extraction, the extracted high-dimensional features first undergo global 
average pooling (GAP). 
This step reduces the number of model parameters while preserving the representative 
information of features across the temporal dimension, thereby mitigating the risk 
of overfitting. 
A dropout layer with a 50\% dropout rate is then inserted between the GAP layer and 
the subsequent fully connected (FC) layers. 
During training, this layer randomly deactivates half of the neurons to prevent the 
model from over-relying on specific features, further suppressing overfitting. 
Following the dropout layer, two FC layers map the pooled features to the final 
classification space and output the probability of the sample belonging to the GRB 
or non-GRB category.
The overall architecture of our ResNet-based model integrated with AFFE modules 
is shown in Figure \ref{fig:network_architectures}, which includes the order of 
layers and the data flow.

Here we describe the AFFE module in detail. Given an input feature 
map $\mathbf{X} \in \mathbb{R}^{T \times C}$ (with $T$ as temporal length and $C$ as 
channel number), the AFFE module processes it in the following steps:

\begin{enumerate}
    \item \textbf{Frequency Transformation}:
    The input feature map $\mathbf{X}$ is first transformed into the frequency 
    domain using the Fast Fourier Transform (FFT):
\begin{equation}
    \mathbf{F} = \text{FFT}(\mathbf{X}),
\end{equation}
    where $\mathbf{F} \in \mathbb{C}^{T \times C}$ 
    denotes the complex-valued 
    Fourier frequency spectrum. 
    The real and imaginary parts of $\mathbf{F} $ are denoted as
    $\mathbf{F}_{\text{real}}$ and $\mathbf{F}_{\text{imag}}$, respectively.

    \item \textbf{Weighted Frequency}:
    To adaptively emphasize important frequency components, we apply separate learnable 
    transformations to the real and imaginary parts of the frequency spectrum. 
    Specifically, four FC layers are used to generate adaptive weights for 
    $\mathbf{F}_{\text{real}}$ and $\mathbf{F}_{\text{imag}}$:
\begin{equation}
    \mathbf{W}_{\text{real}} = \text{FC}_{2}(\text{FC}_{1}(\mathbf{F}_{\text{real}})), \quad \mathbf{W}_{\text{imag}} = \text{FC}_{4}(\text{FC}_{3}(\mathbf{F}_{\text{imag}})),
\end{equation}  
    where $\mathbf{W}_{\text{real}}, \mathbf{W}_{\text{imag}} \in \mathbb{R}^{T \times C}$ 
    are the learned weight matrices. The number of neurons in the $\text{FC}_{1}$ and 
    $\text{FC}_{3}$ is set to $C$ divided by 8. 
    This choice is determined after comparing division factors of 2, 4, 8, and 16, 
    and it significantly reduces the parameter count of the module. 
    The weighted frequency spectrum is then computed as:
\begin{equation}
    \mathbf{F}_{\text{weighted}} = \mathbf{W}_{\text{real}} \odot \mathbf{F}_{\text{real}} + i \cdot (\mathbf{W}_{\text{imag}} \odot \mathbf{F}_{\text{imag}}),
\end{equation}
    where $\odot$ denotes element-wise multiplication.

    \item \textbf{Filtered Frequency}:
    To further refine frequency features, we perform a filtering operation in the 
    frequency domain. 
    First, the weighted frequency spectrum $\mathbf{F}_{\text{weighted}}$ is summed 
    along the channel dimension to obtain a channel-aggregated frequency 
    representation $\mathbf{F}_{\text{sum}} \in \mathbb{C}^{T \times 1}$. 
    Two one-dimensional convolutional layers are then applied to $\mathbf{F}_{\text{sum}}$ 
    to learn a frequency thresholding function that approximates a low-pass filter:
\begin{equation}
    \mathbf{F}_{\text{filtered}} = \text{Conv}_1(\text{Conv}_2(\mathbf{F}_{\text{sum}})).
\end{equation}
    The filtered spectrum $\mathbf{F}_{\text{filtered}}$ is then broadcasted 
    back to the original channel dimension.

    \item \textbf{Frequency Feature Fusion}:
    The results of the Weighted Frequency and Filtered Frequency steps are 
    combined through element-wise addition:
\begin{equation}\label{f_enhanced}
    \mathbf{F}_{\text{enhanced}} = \mathbf{F} \odot \mathbf{F}_{\text{weighted}} + \mathbf{F}  \odot \mathbf{F}_{\text{filtered}}.
\end{equation}

    \item \textbf{Inverse Transformation}:
    Finally, the enhanced frequency spectrum $\mathbf{F}_{\text{enhanced}}$ 
    is transformed back to the time domain using the Inverse Fast Fourier Transform (IFFT):
\begin{equation}
    \mathbf{X}_{\text{enhanced}} = \text{IFFT}(\mathbf{F}_{\text{enhanced}}),
\end{equation}
    where $\mathbf{X}_{\text{enhanced}} \in \mathbb{R}^{T \times C}$ is the 
    output feature map with enriched frequency-domain information.
\end{enumerate}

\subsection{Model training and analysis}
\label{sec:train_model}

Model parameters are initialized using the truncated normal distribution 
proposed by \citet{param_he_normal}. 
A detailed analysis of hyperparameter choices-including batch size, learning 
rate, and number of convolutional layers is conducted to evaluate their impact 
on the test set, following the approach in \citet{dl_technosignatures}. 
Additional hyperparameters, including the number of convolutional kernels per unit, the size of the convolutional kernels, and the neuron counts in the fully connected (FC) layers, were also optimized.
The training configuration is as follows: batch size is set to 1024; cross-entropy 
loss is used to measure the discrepancy between predicted and true labels; 
and the Adam optimizer \citet{adam_optimiser} is adopted to dynamically update 
model parameters and minimize loss during training. 
The initial learning rate is $1 \times 10^{-4}$, with a scheduling mechanism 
that reduces the rate by half if validation loss does not decrease for 10 
consecutive epochs. 
To prevent overfitting, early stopping terminates training if validation accuracy 
plateaus for 20 consecutive epochs.
The patience value was selected after testing ranges of [10, 20, 40] epochs, 
and the model parameters from the epoch with the highest validation accuracy are 
retained as the final optimal parameters.
The result of the optimizing choices are highlighted in bold in 
Table \ref{table 2:hyper_parameter_select}. 
The efficacy of selected hyperparameters is further validated by training curves, 
which demonstrate stable convergence and appropriate model capacity for 
the classification task.

\begin{deluxetable*}{cc}[htbp]
\label{table 2:hyper_parameter_select}
\tablecaption{Hyper-parameter selection.}
\tablehead{
Parameters & Values}
\startdata
Number of ConvUnit & 2, \textbf{4}, 8, 16 \\
Number of Conv in ConvUnit & 1, 2, \textbf{3}, 4 \\
First ConvUnit filter size & 16, \textbf{32}, 64, 128, 256, 512 \\
Second ConvUnit filter size & 16, \textbf{32}, 64, 128, 256, 512 \\
Third ConvUnit filter size & 16, 32, \textbf{64}, 128, 256, 512 \\
Fourth ConvUnit filter size & 16, 32, \textbf{64}, 128, 256, 512 \\
Norm function after Conv & InstaceNorm, \textbf{BatchNorm} \\
Activation function & Sigmoid, \textbf{Relu} \\
FC neurons & 8, 16, 32, \textbf{64}, 128 \\
Dropout rate & 0.3, \textbf{0.5}, 0.8 \\
\hline\noalign{\smallskip}
Initial learning rate & 1e-3, \textbf{1e-4}, 1e-5, 1e-6 \\
Batch size & 512, \textbf{1024}, 2048\\ 
Patience of reduce learning rate & 5, \textbf{10}, 15\\
\enddata
\tablecomments{Bold text represents that the model performs optimally 
on that metric.}
\end{deluxetable*}

We conducted a comparative analysis involving the baseline 
ResNet architecture that is commonly used for time series data, and ResNet variants 
integrated with existing feature enhancement modules that including the SE 
(Squeeze-and-Excitation block; \citet{model_block_SE}), 
CBAM (Convolutional Block Attention Module; \citet{model_block_CBAM}), 
ECA (Efficient Channel Attention module; \citet{model_block_ECA}), 
and ASB (Adaptive Spectral Block; \citet{model_block_ASB} modules), 
and our proposed ResNet-AFFE model.
All models were trained and tested on our dataset using hyperparameters consistent 
with those reported in the original literature for each comparative module. 
This ensures a fair performance comparison, and the results of this comparison 
(for ResNet variants and our ResNet-AFFE model) on the test set are 
presented below.

The models were implemented using the PyTorch framework, with training 
conducted on a single NVIDIA RTX-4090 GPU. 
Performance was evaluated using four standard classification metrics: 
\textit{Accuracy}, \textit{Precision}, \textit{Recall}, and \textit{F1-score}, 
consistent with the evaluation methodology reported 
in \citet{DL_identify_grb_by_peng}.
Accurately quantifying these metrics and their associated 
prediction uncertainties is critical, as they quantify confidence in model 
predictions and provide a scientific basis for subsequent decision-making. 
During testing, we achieved robust uncertainty quantification for deep learning 
predictions via Monte Carlo Dropout (MCD, 
for references see \citet{model_uncertainty1,model_uncertainty2}). 
In this procedure, all model parameters are kept fixed, and dropout layers 
between FC layers remain activated. For each forward pass, the dropout rate 
is randomly sampled from a uniform distribution over the interval [0.1, 0.5].
This stochastic forward pass is repeated 1000 times, which allows computation 
of the mean and standard deviation for each metric—thereby effectively 
estimating predictive uncertainty. The performance and uncertainty of all models
are summarized in Table~\ref{table3}.

\begin{deluxetable*}{ccccc}[htbp]
\label{table3}
\tablecaption{Comparison of models' performance on test set.}
\tablehead{
Model & \textit{Accuracy} (\%) & \textit{Precision} (\%) & \textit{Recall} (\%) & \textit{F1-score} (\%)}
\startdata
ResNet & 97.17 $\pm$ 0.13 & 98.80 $\pm$ 0.14 & 94.71 $\pm$ 0.20 & 96.72 $\pm$ 0.15 \\
ResNet+SE$^a$ & 97.33 $\pm$ 0.04 & 99.08 $\pm$ 0.05 & 94.82 $\pm$ 0.08 & 96.90 $\pm$ 0.04 \\
ResNet+CBAM$^b$ & 97.29 $\pm$ 0.07 & 99.08 $\pm$ 0.08 & 94.72 $\pm$ 0.14 & 96.85 $\pm$ 0.09 \\
ResNet+ECA$^c$ & 97.34 $\pm$ 0.07 & 99.15 $\pm$ 0.08 & 94.76 $\pm$ 0.13 & 96.91 $\pm$ 0.08 \\
ResNet+ASB$^d$ & 97.40 $\pm$ 0.07 & 99.24 $\pm$ 0.08 & \textbf{94.83 $\pm$ 0.14} & 96.98 $\pm$ 0.09 \\
ResNet+AFFE (Ours) & \textbf{97.46 $\pm$ 0.07} & \textbf{99.39 $\pm$ 0.07} & 94.81 $\pm$ 0.14 & \textbf{97.04 $\pm$ 0.08} \\
\enddata
\tablecomments{
The uncertainty of each metric is evaluated utilizing MCD.
Bold text represents that the model performs optimally on that metric.\\
Enhanced time domain feature extraction module: $^a$~\citet{model_block_SE}, $^b$~\citet{model_block_CBAM}, $^c$~\citet{model_block_ECA}.\\
Enhanced frequency domain feature extraction module: $^d$~\citet{model_block_ASB}.
}
\end{deluxetable*}

\begin{figure}
\centering
\subfigure{
\includegraphics[width=0.8\linewidth]{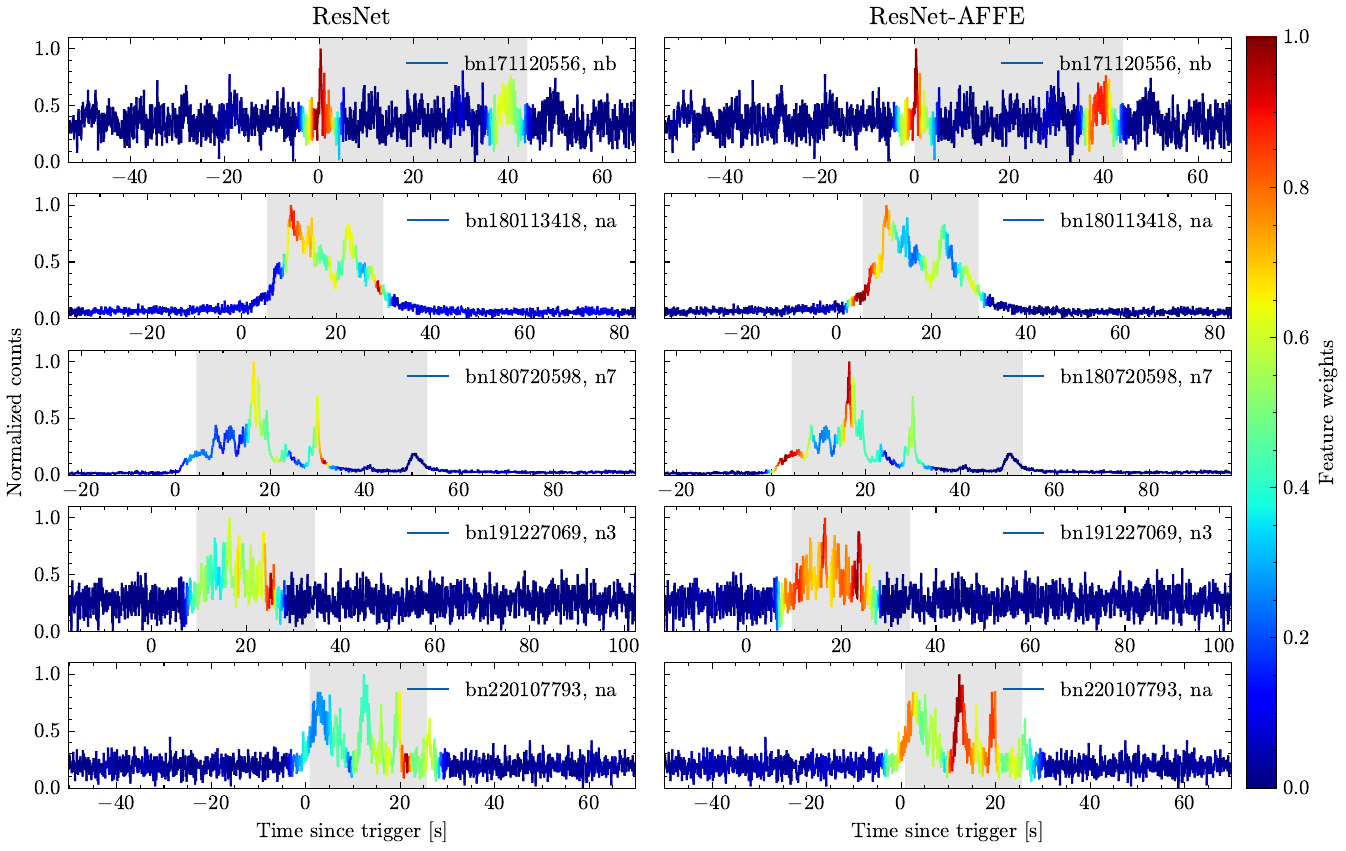}
}
\caption{Feature visualization for five representative GRBs using Grad-CAM. 
The color-coded heatmaps indicate the temporal regions receiving strongest attention 
from the deep learning model during classification. 
Left and right panels contrast the feature extraction patterns between the 
ResNet (left) and our enhanced ResNet-AFFE architecture (right). 
Gray shaded regions denote the $T_{90}$ for each GRB.}
\label{fig:feature_cam_compare}
\end{figure}

To examine whether ResNet-AFFE prioritizes burst-specific temporal signatures 
(e.g., $T_{90}$ intervals or energy-dependent flux peaks) over noise, 
we used Gradient-weighted Class Activation Mapping (Grad-CAM)—a mainstream visualization 
technique for addressing deep learning’s "black box" problem. 
Grad-CAM generates gradient-weighted heatmaps to identify features most 
influential to model predictions \citep{model_grad_cam}, enhancing interpretability. 
For time-series light curves, these heatmaps highlight temporal 
segments driving classification outputs, replacing ambiguous "feature importance" 
with intuitive visual evidence. 
We employ Grad-CAM to decode our model's decision-making process for GRB classification. 
Comparative visualizations of discriminative temporal features between the 
baseline ResNet and ResNet-AFFE are presented in Figure \ref{fig:feature_cam_compare}.

\begin{figure*}[htbp]
\centering
\subfigure{
\includegraphics[width=0.45\linewidth]{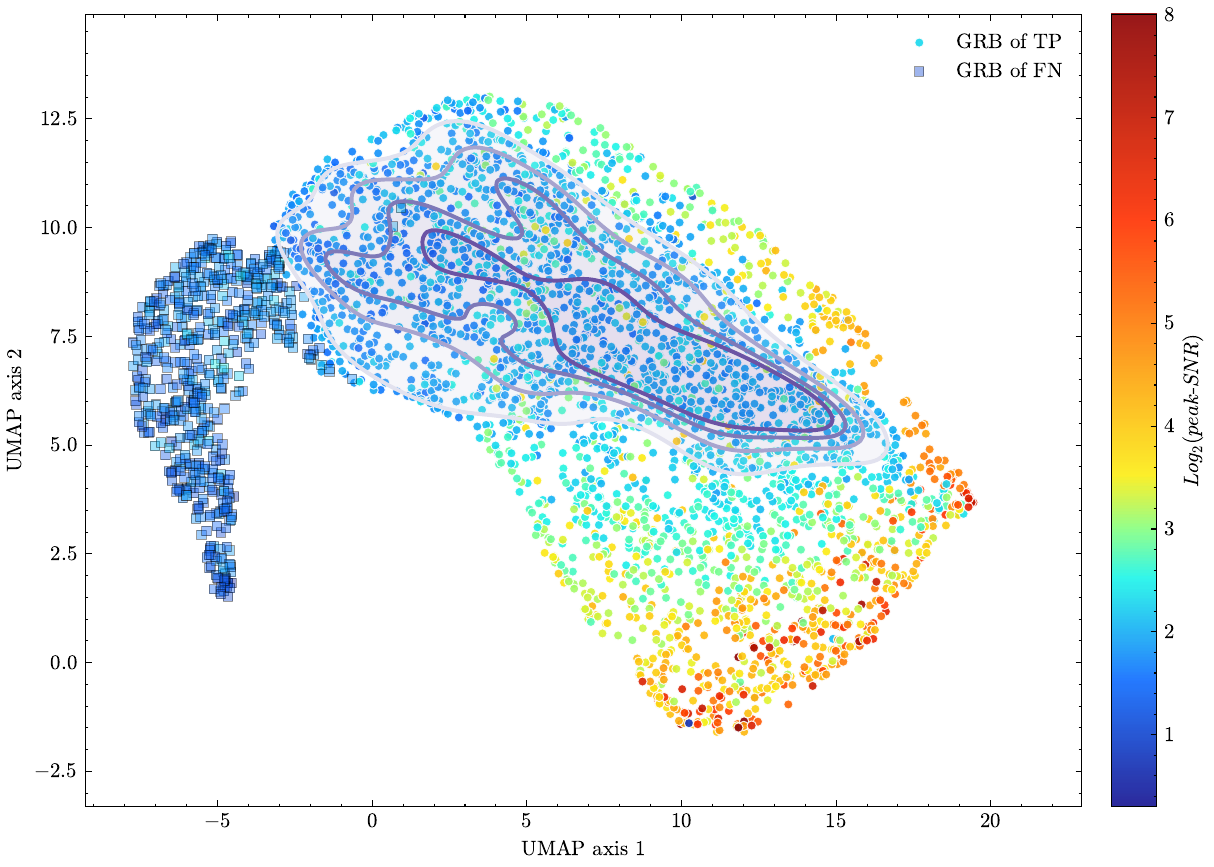}
}
\subfigure{
\includegraphics[width=0.45\linewidth]{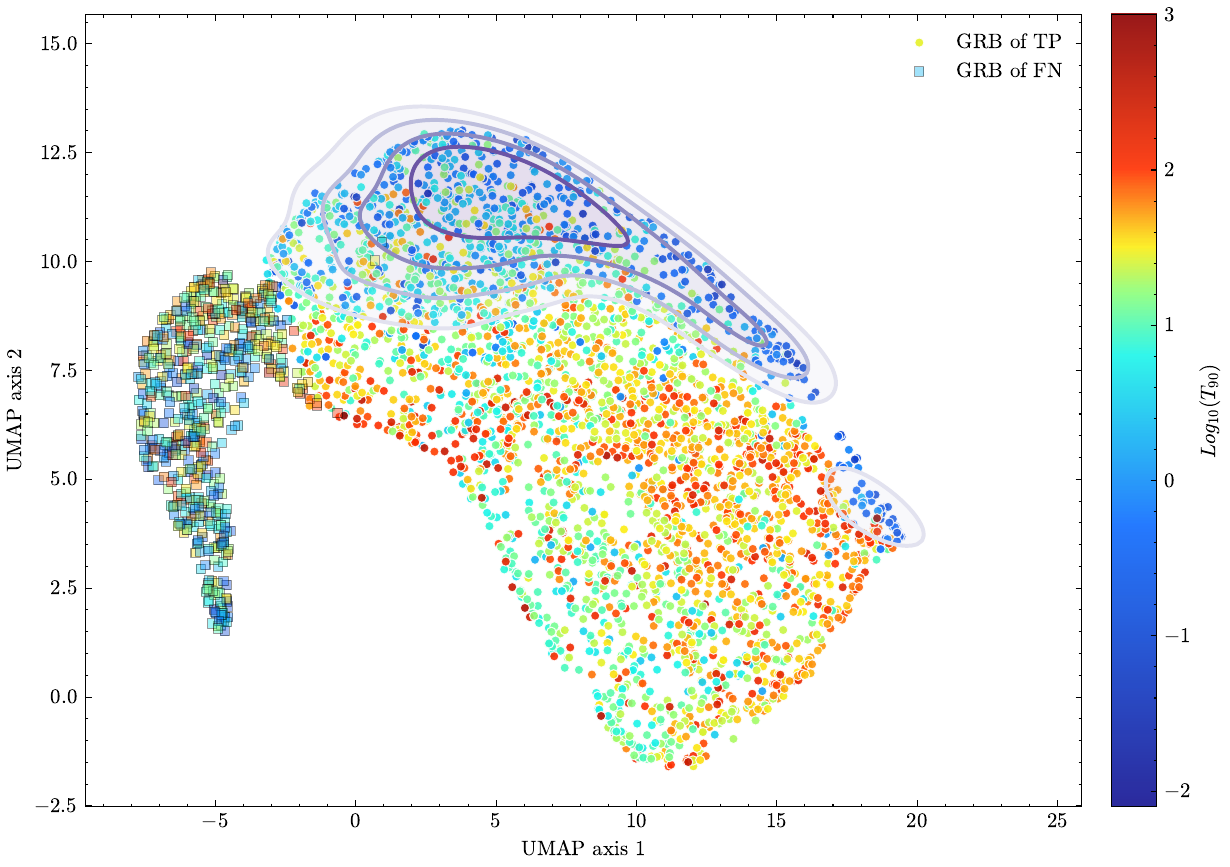}
}
\caption{
UMAP visualization of feature space representations extracted from the final 
convolutional layer of the ResNet-AFFE model.
(Left) Points are color-coded according to $\log_{2}(\mathrm{peak\mbox{-}SNR})$ 
values for individual GRBs.
The purple contours represent the kernel density estimate (KDE) of low-significance 
TP events ($\mathrm{peak\mbox{-}SNR} < 5\sigma$).
(Right) Points are color-coded according to $\log_{10}(T_{90})$ values obtained 
from the \textit{Fermi} burst catalog.
Similarly, the purple contours delineate the high-density region of short-duration TP events ($T_{90} < 2\,\mathrm{s}$).
}
\label{fig:down_dim}
\end{figure*}

We further mapped deep features extracted by ResNet-AFFE to observed physical 
characteristics using UMAP, a nonlinear dimensionality reduction method rooted 
in graph theory and manifold learning that effectively projects high-dimensional 
data to low-dimensional spaces for structural analysis.
We analyzed 3,833 GRBs detected by Fermi/GBM over the period 2008–2024, 
concatenating light curves from the triggering detectors of each GRB as model input. 
Notably, our model uses GAP to aggregate features from the last convolutional layer, 
enabling compatibility with variable-length data under uniform batch input. 
This design specifically supports the inclusion of long GRBs with $T_{90} > 120\ \text{s}$, 
which would otherwise be excluded by fixed-length constraints.
UMAP takes as input the output results from the last convolutional layer of 
ResNet-AFFE and outputs 2D projections that uncover intrinsic correlations of features
between GRBs. 
Key UMAP hyperparameters including \textit{n\_neighbors} and \textit{min\_dist} were 
systematically optimized via exhaustive parameter space exploration (over ranges: 
$2 \leq \textit{n\_neighbors} \leq 100$; $0.01 \leq \textit{min\_dist} \leq 0.99$). 
Manual validation yielded optimal values of  $\textit{n\_neighbors} = 60$ and 
$\textit{min\_dist} = 0.66$, achieving a balanced preservation of both global data 
structure and local neighborhood fidelity—two core advantages of UMAP in 
high-dimensional signal analysis.
The resulting projections, shown in Figure \ref{fig:down_dim}, visualize the distribution 
of true positive (TP) and false negative (FN) GRBs, with each GRB’s peak-SNR and $T_{90}$ 
encoded directly in the projection to link low-dimensional structure with physical 
observables.
While UMAP effectively reveals the structure of high-dimensional data, 
its visualization results are sensitive to hyperparameter selection 
(e.g., $\textit{n\_neighbors}$ and $\textit{min\_dist}$). 
To mitigate the subjectivity introduced by manual parameter tuning and ensure the 
robustness of observed clustering patterns, we implemented a cross-validation procedure.

\begin{figure}
\centering
\includegraphics[width=\textwidth, angle=0]{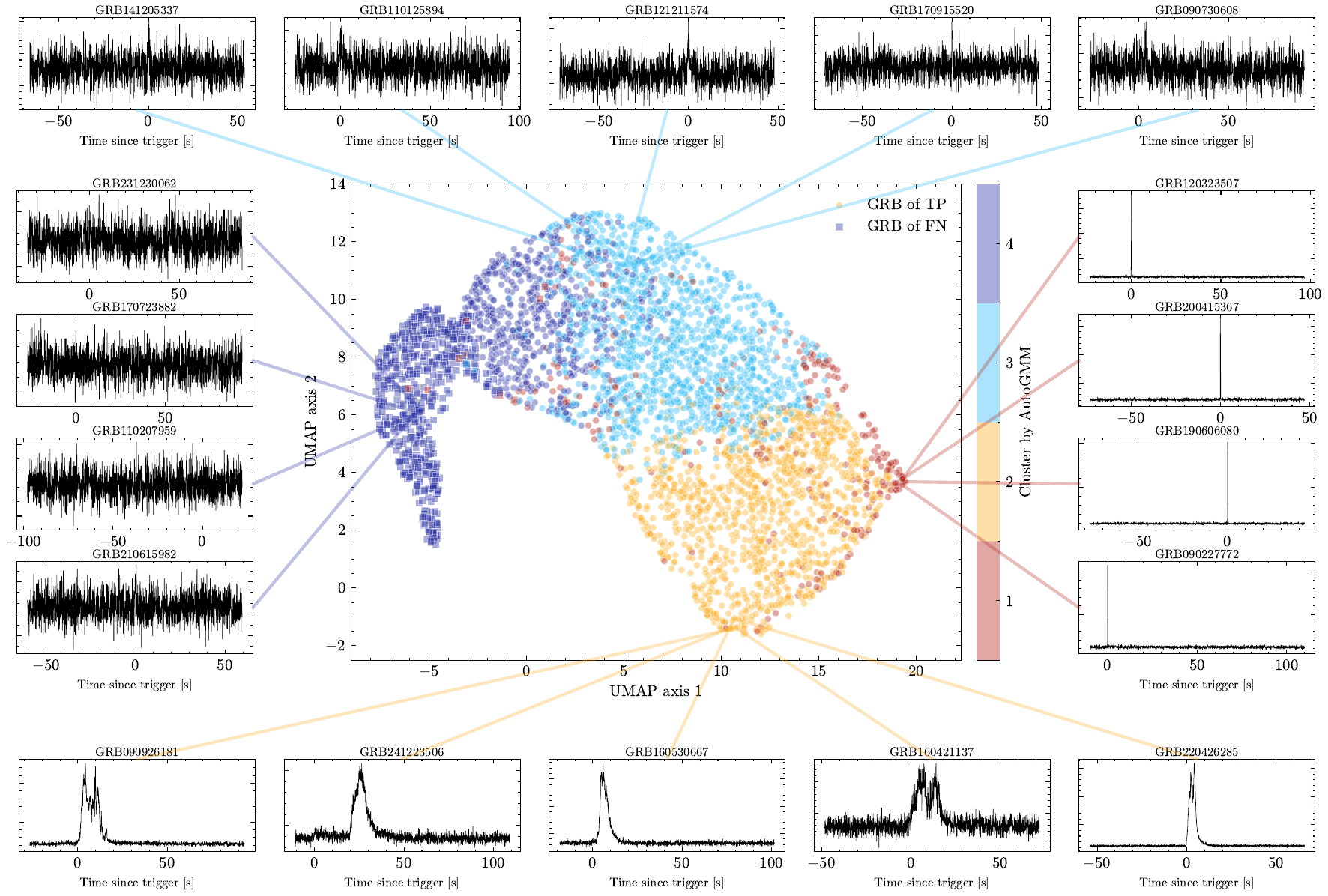}
\caption{The dimensionality reduction results of UMAP algorithm for the output features of 
the last convolution layer of the ResNet-AFFE model, colored by their AutoGMM 
cluster assignments (four clusters). 
Surrounding panels display the full-band light curves of representative GRBs 
from each cluster.
}
\label{fig:down_dim_with_lc_example}
\end{figure}

\begin{figure*}[htbp]
\begin{overpic}[width=0.5\textwidth]{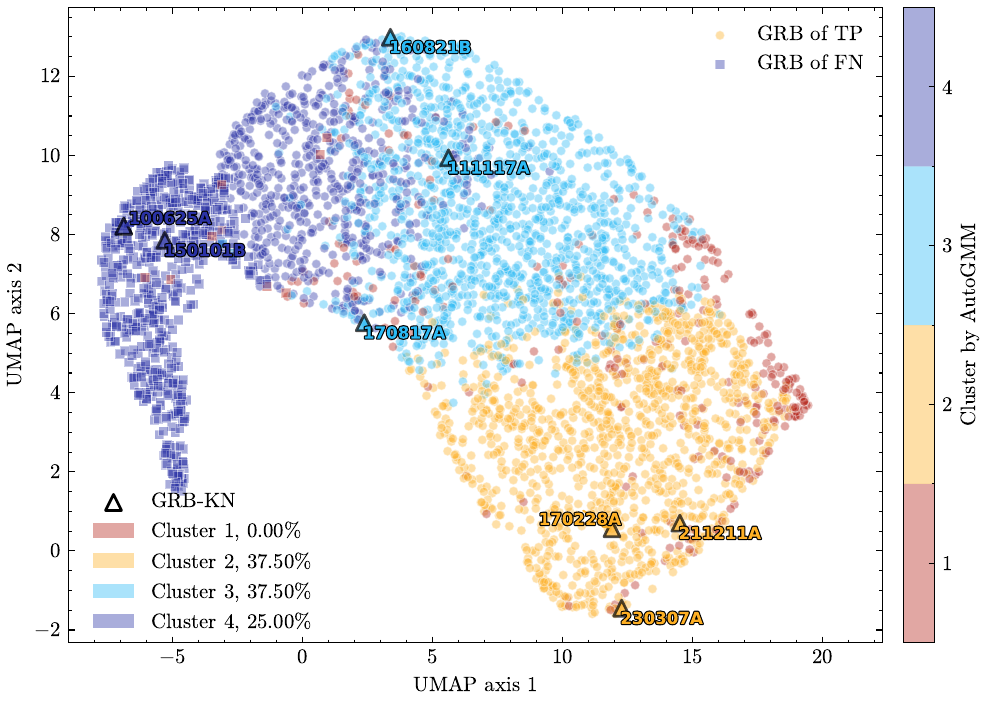}
    \put(-1,65){(a)}
\end{overpic}
\hspace{2mm}
\begin{overpic}[width=0.5\textwidth]{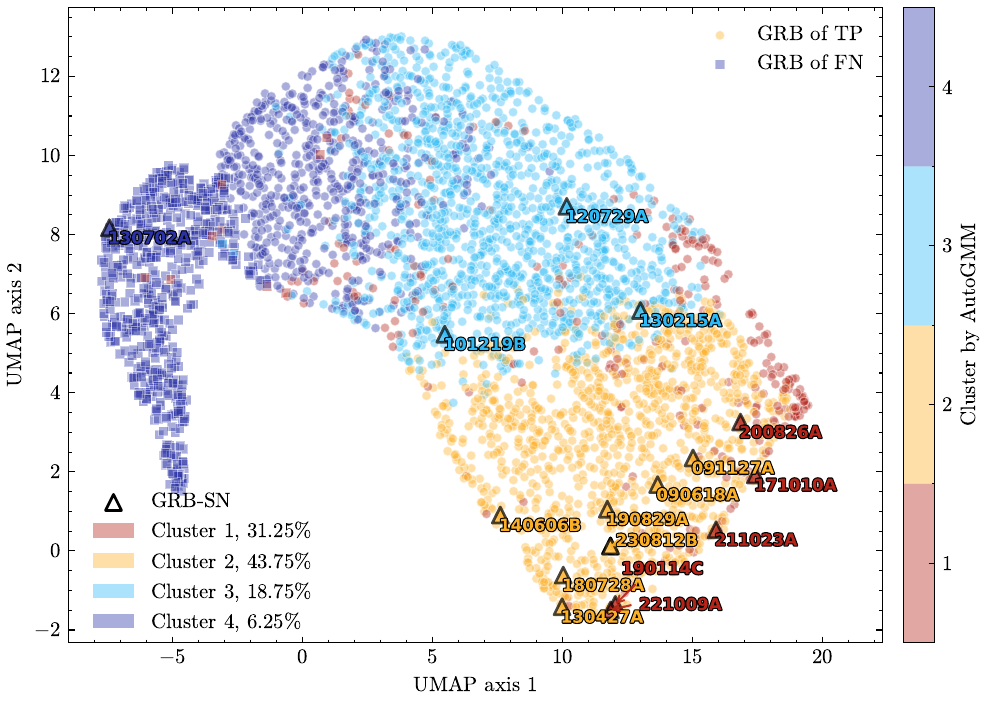}
    \put(-1,65){(b)}
\end{overpic}
\hspace{2mm}
\begin{overpic}[width=0.5\textwidth]{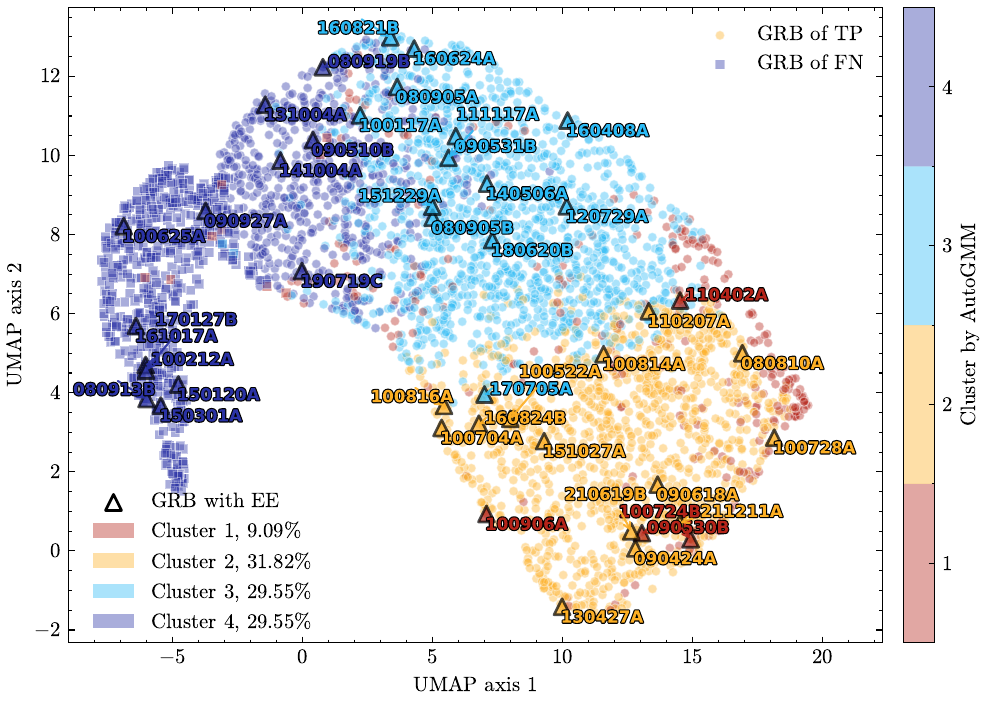}
    \put(-1,65){(c)}
\end{overpic}
\hspace{2mm}
\begin{overpic}[width=0.5\textwidth]{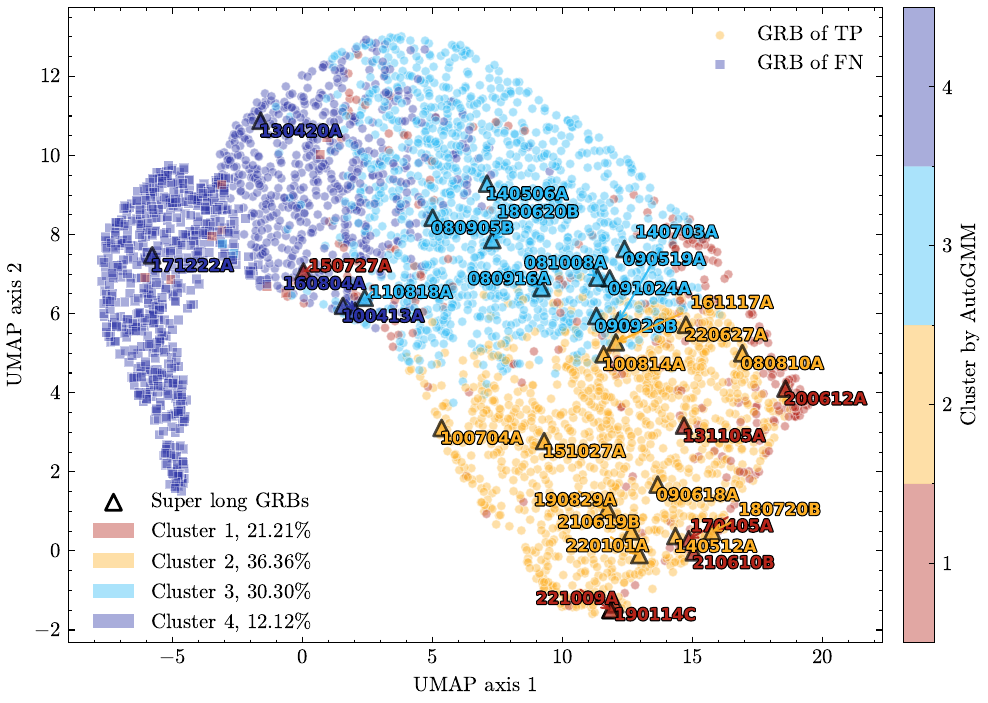}
    \put(-1,65){(d)}
\end{overpic}
\hspace{2mm}
\begin{overpic}[width=0.5\textwidth]{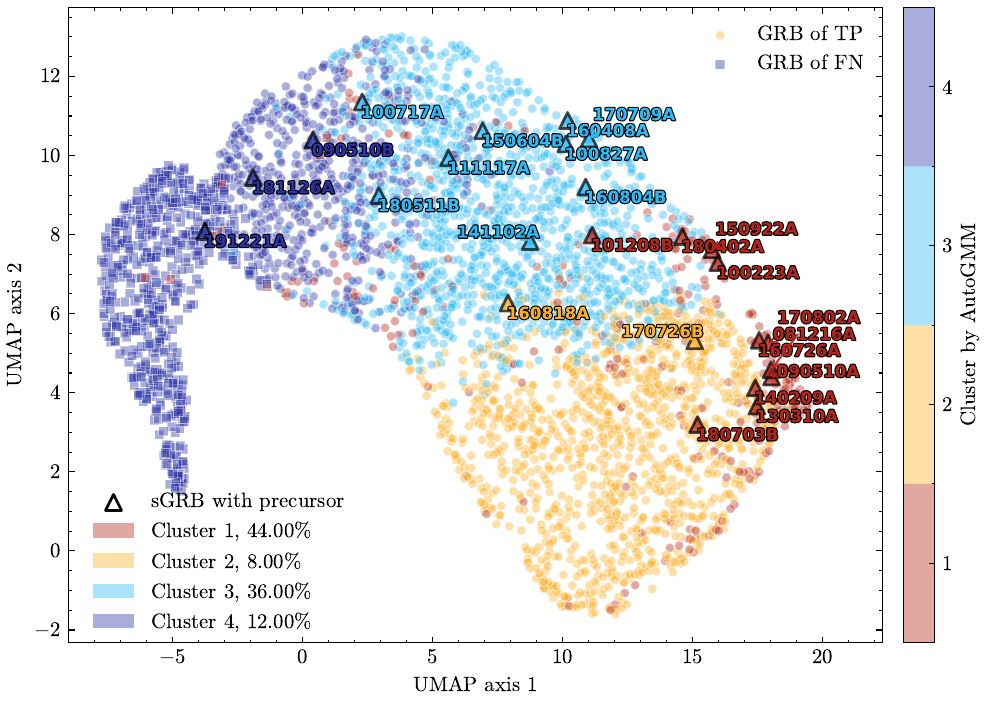}
    \put(-1,65){(e)}
\end{overpic}
\hspace{2mm}
\begin{overpic}[width=0.5\textwidth]{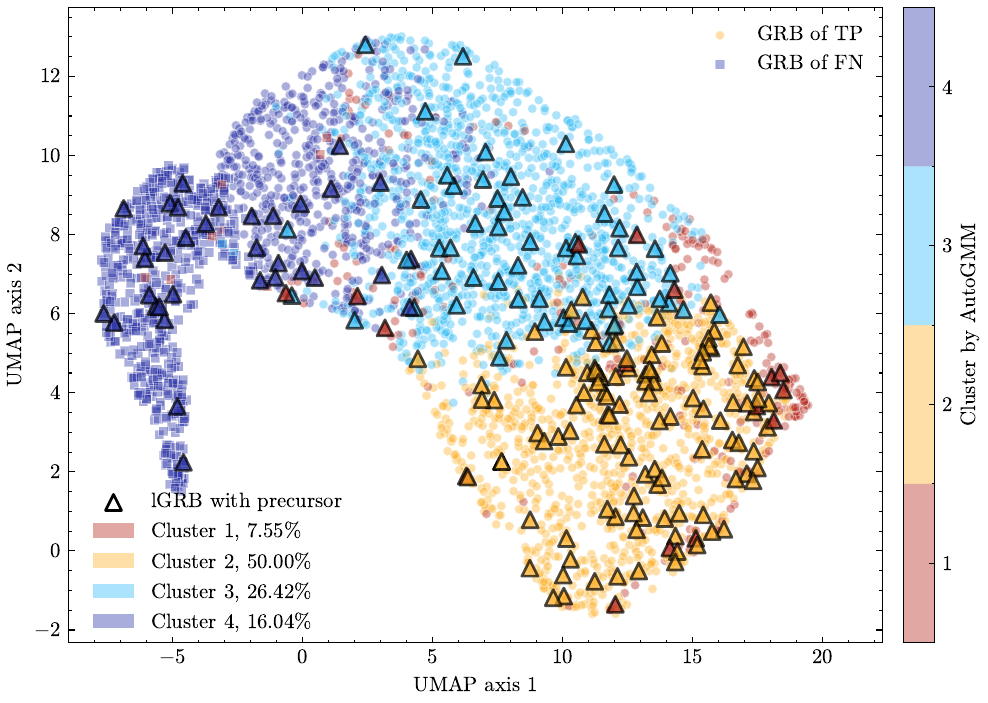}
    \put(-1,65){(f)}
\end{overpic}
\caption{The dimensionality reduction results of UMAP algorithm for the output features of 
the last convolution layer of the ResNet-AFFE model, colored by their AutoGMM cluster 
assignments (four clusters).
The triangular symbols in each sub-graph are the same series of GRBs detected by Fermi/GBM: 
$(a)$ Kilonova-associated GRBs (GRB-KN, \cite{GRB-KN_2024Li}),
$(b)$ Supernova-associated GRBs (GRB-SN) come from the GRBSN webtool~\citep{GRB-SN_2024Finneran}, 
$(c)$ GRBs with extended emission (GRB with EE, \cite{ML-downdim-GRB-EE_2023Garcia, GRB-EE_2024Li}),  
$(d)$ Super long GRBs~\citep{GRB-superlong_2024Ror}, 
$(e)$ Short GRBs with precursor~\citep{sGRB-precursor_2019Zhong,sGRB-precursor_2020Wang}, 
$(f)$ Long GRBs with precursor~\citep{lGRB-precursor_2020Coppin,lGRB-precursor_2024Deng}. 
}
\label{fig:down_dim_special_grb}
\end{figure*}

Following the approach in \citet{DownDimGRB_Dimple2023}, which targets 
latent clustering patterns in transient burst events, we further employed the AutoGMM 
module \citep{autoGMM} to cluster features extracted by the ResNet-AFFE model. 
AutoGMM is an automated clustering tool that leverages Gaussian Mixture Models (GMMs) 
to infer the optimal number of clusters, and it assumes data points arise from a mixture 
of multiple Gaussian distributions, aligning with the statistical properties of GRB 
feature distributions. 
The algorithm first performs GMM clustering over a range of candidate cluster counts, 
then infers the Gaussian distribution parameters and cluster assignments that best fit 
the data—eliminating subjective manual cluster number selection.
The four clusters identified via AutoGMM were visually distinguished using distinct 
color mappings in the UMAP projections, and their corresponding representative light 
curves were presented around, as shown in Figure \ref{fig:down_dim_with_lc_example}. 
Figure \ref{fig:down_dim_special_grb} further delineates the spatial distribution 
of physically distinct GRB subgroups and their proportions within each cluster, 
including kilonova-associated GRBs (GRB-KN), supernova-associated GRBs (GRB-SN), 
extended-emission GRBs (GRB-EE), ultra-long GRBs, 
and short/long GRBs with precursors.

\section{Result}
\label{sec:result}

All datasets used in this study are derived from Fermi/GBM observational data, 
divided into three defined subsets for training, validation, and testing. 
The training set incorporates synthetic GRB samples generated via our proposed 
data augmentation technique, designed to increase sample size and enhance the 
diversity of burst signals. 
The efficacy of this method is visually validated in Figure \ref{fig:crop_example}, 
which illustrates controlled modulation of burst SNR while preserving background 
levels. 
Focused on sampling and attenuation factors, the augmentation notably enriches 
low-SNR GRB representation: as shown in Figure \ref{fig:trainset_snr_hist}, 
the peak-SNR distribution of GRB samples expands toward lower values, 
increasing total GRB training samples from 6,189 to over 100,000. 
Non-GRB samples, sourced from triggerless time-tagged event data, were 
similarly augmented to 100,000, ensuring class balance. 
For validation and testing, we used 2,774 and 3,143 primary GRBs, 
respectively, selected based on the number of triggering NaI detectors. 
Detailed dataset specifications are provided in Table \ref{table:dataset}.
Each input sample consists of light curves across 9 energy bands 
(64 ms temporal resolution; Figure \ref{fig:network_architectures}, left), 
with per-band standardization applied to reduce energy-specific magnitude biases. 
This preprocessing preserves fine-grained features, enhanced the model’s 
feature extraction and generalization capabilities, 
and facilitated subsequent visualization analyses.

Our proposed classification framework employs a ResNet-based 1D-CNN within 
a supervised learning paradigm, designed for binary classification 
(GRB vs. non-GRB). 
Its core capability lies in extracting discriminative features from 
multidimensional data and differentiating classes based on these features. 
The complete model architecture—incorporating the novel AFFE module—is shown 
in Figure \ref{fig:network_architectures}.
The AFFE module employs a dual-path filtering architecture 
(Equation \ref{f_enhanced}): the term $\mathbf{F} \odot \mathbf{F}_{\text{weighted}}$
performs frequency-adaptive soft-threshold denoising, 
while  $\mathbf{F} \odot \mathbf{F}_{\text{filtered}}$ implements coherent 
feature extraction via an optimized matched-filter scheme. 
This integrated approach enables adaptive frequency selection without 
manual filter design, achieving simultaneous noise suppression and signal 
enhancement through data-driven spectral weighting. 
By weighting and filtering frequency-domain differences between GRBs and non-GRBs, 
the module optimizes extraction of discriminative features critical for classification.

Model parameters were optimized via backpropagation and gradient-based methods 
on our large-scale datasets, enhancing autonomous learning of salient features. 
Systematic evaluation of hyperparameter configurations identified the optimal 
set (Table \ref{table 2:hyper_parameter_select}). 
Training and validation loss curves exhibit stable dynamics: 
the early stopping mechanism terminated training at epoch 18 (the optimal point), 
preserving parameters that generalize best to unseen data. 
Convergence of both curves—with no divergence between training and validation 
losses—confirms robust performance without overfitting.
We conducted a comparative analysis of multiple models on the test set, 
including: a baseline ResNet, ResNet variants with existing feature enhancement 
modules (SE, CBAM, ECA, ASB), and our ResNet-AFFE. 
All models were trained with hyperparameters consistent with their original 
literature to ensure fairness, and prediction confidence was estimated via 
Monte Carlo Dropout (1000 stochastic forward passes). 
Performance metrics and uncertainties are summarized in Table \ref{table3}.
Incorporating the AFFE module yielded a statistically quantifiable improvement 
in classification accuracy. 
Specifically, ResNet-AFFE achieved 97.46\% accuracy—outperforming all tested 
variants with conventional enhancement modules—with consistent results across 
multiple experimental runs.
Grad-CAM visualized the model’s decision-making process, revealing that 
ResNet-AFFE prioritizes physically meaningful GRB characteristics. 
As shown in Figure \ref{fig:feature_cam_compare}, compared to the baseline ResNet, 
our model’s attention is more strongly concentrated in the \(T_{90}\) interval 
(containing key prompt emission information) rather than random noise—highlighting 
a critical difference in feature focus. 
This visualization enables even untrained users to distinguish model robustness, 
even for identical predictions.

We applied UMAP to reduce the dimensionality of features from the last 
convolutional layer of ResNet-AFFE, using 3,833 original GRBs as input. 
Figure \ref{fig:down_dim} presents the 2D projection, encoded with individual 
GRBs’ peak-SNR and \(T_{90}\) values, revealing distinct clustering patterns: 
true positive (TP) and false negative (FN) GRBs occupy entirely separate regions. 
Of these GRBs, 593 were classified as FNs—99\% with peak-SNR below 5\(\sigma\) 
and most being short-duration events.
Dimensionality reduction further uncovered organized spatial distributions: 
low-SNR GRBs clustered in the upper-left region, while short-duration GRBs 
formed a narrow band along the upper-right edge, exhibiting a characteristic 
nonlinear distribution. Notably, GRBs with similar durations or SNRs showed 
discernible aggregation.
Automated clustering via AutoGMM identified four distinct clusters from the 
features extracted by the ResNet-AFFE model eliminating subjective manual 
selection. 
Figure \ref{fig:down_dim_with_lc_example} illustrates representative light 
curves for each cluster, mapped onto the UMAP-reduced space, directly linking 
clustering results to observable temporal morphology of GRBs.
To further explore the physical relevance of clustering, we analyzed 
correlations between model-extracted features and GRBs of specific origins 
or morphologies, quantifying the proportion of these GRBs in each cluster 
to preliminarily assess associations. 
Figure \ref{fig:down_dim_special_grb} highlights these special GRBs in the 
dimensionality reduction projection:
Kilonova-associated GRBs are predominantly localized within Clusters 2 and 3, 
while supernova-associated events are primarily concentrated in Clusters 1 and 2. 
GRBs with extended emission are uniformly distributed across Clusters 2 to 4. 
Ultra-long GRBs are spatially confined to the lGRB-dominated Cluster 3, 
suggesting distinct feature space signatures. 
Short GRBs with precursors exhibit a concentrated spatial distribution within 
Cluster 1, occupying both peripheral regions and transitional zones bridging 
short and long GRB populations. 
Long GRBs with precursors contrastingly show half their population concentrated 
in Cluster 2 with uniform dispersion.
This visualization reveals natural separations in the learned feature space between GRBs of different physical origins.

\section{Discussion and Conclusion}
\label{sec:discussion}

This study develops an integrated deep learning framework for high-precision 
GRB identification, addressing core challenges in transient astrophysical 
event detection through targeted methodological innovations. 
The transient nature of GRBs poses a fundamental obstacle to deep learning 
applications, marked by limited observable events and significant observational 
bias favoring high-significance bursts—two longstanding limitations that 
our physics-informed data augmentation framework directly overcomes.

By generating over 100,000 synthetic GRB samples that an order-of-magnitude 
expansion from the original 6,189 real-world observations, 
this framework effectively mitigates observational selection bias against 
low-significance events and helps to address the "observational iceberg" effect
described by \cite{2014MNRAS.442.1922L} and \cite{iceberg-effects_2022Moss}, 
where the majority of astrophysical transients lie below conventional 
detection thresholds due to instrumental sensitivity limitations and 
background contamination.

As demonstrated in Figure \ref{fig:trainset_snr_hist}, our data augmentation methodology successfully 
reduces the observational bias against low-significance events, 
which constitute the predominant population of actual GRB phenomena yet 
remain systematically underrepresented in existing catalogs. 
Specifically, we bin light curves into 9 distinct energy channels following 
traditional trigger search 
algorithms \citep{fermi_gbm_fouth_catalog, search_grb_hxmt_cai}, 
preserving crucial spectral information while enhancing signal detectability. 
We then systematically modulate burst SNR while retaining realistic background 
characteristics, yielding physically faithful synthetic samples that span the 
full dynamic range of actual GRB phenomena. 
This approach not only reduces model overfitting but also enhances sensitivity 
to weak transients—critical for capturing the majority of GRBs that lie below 
conventional detection thresholds due to instrumental constraints. 
The proposed augmentation framework offers substantial advantages for GRB 
identification systems by alleviating several fundamental limitations 
in current approaches. 
Furthermore, the enhanced sample diversity allows the development of detection 
algorithms with improved sensitivity to weak transients that typically evade
identification in standard analysis pipelines.
As such, the augmentation strategy provides a generalizable solution for 
transient astronomy, where limited mission durations and observational biases 
often hinder deep learning applications.

We design a novel Adaptive Frequency Feature Enhancement module, AFFE, which 
represents a methodological advance in addressing the spectral-temporal 
complexity of GRB signals. 
Unlike conventional time-domain attention mechanisms (e.g., SE, CBAM, ECA) 
that fail to capture frequency-domain correlations, or fixed-frequency filters 
lacking adaptability, AFFE employs a dual-path architecture to perform 
data-driven spectral weighting—enabling adaptive soft-threshold denoising 
and coherent feature extraction without manual filter design. 
This mechanism aligns with the complex, variable frequency signatures 
of GRBs, optimizing the model’s ability to distinguish GRBs from non-GRB 
signals by leveraging discriminative frequency components. 
Integrating AFFE with a ResNet baseline yields a classification accuracy of 
97.46\%, representing a statistically significant 3\% improvement over the 
state-of-the-art 2D ResNet-CBAM architecture (94.46\%; \citep{DL_identify_grb_by_peng}). 
Comparative analysis of these models and their predictive uncertainty reveal 
that our AFFE implementation shows superior performance metrics, 
establishing a new benchmark for high-precision GRB identification.
Comparative analysis further validates that this frequency-adaptive 
architecture maintains performance advantages even near theoretical accuracy 
saturation, demonstrating its superiority and robustness in handling the 
unique spectral dynamics of GRB light curves.

Further validation confirms that this frequency-adaptive architecture 
maintains performance advantages even near theoretical accuracy saturation, 
demonstrating its superiority and robustness in handling the unique spectral 
dynamics of GRB light curves, highlighting the methodological advancement 
in alleviating the unique challenges of GRB identification. 
As visualized in Figure \ref{fig:feature_cam_compare}, 
the model selectively attends to critical burst characteristics, 
exhibiting strong concordance with domain expertise in GRB identification.
The incorporation of the AFFE module significantly enhances the ResNet 
architecture’s capacity to extract physically meaningful features, 
enabling comprehensive characterization of burst processes. 
This Fourier-domain enhancement improves classification by explicitly 
modeling spectral-temporal correlations—critical for distinguishing GRBs 
from noise, as key burst signatures (e.g., multi-phase emission, spectral lags) 
are jointly encoded in both time and frequency domains.

Additionally, Grad-CAM feature visualization and UMAP dimensionality reduction 
confirm the model’s ability to focus on physically meaningful \(T_{90}\) regions 
and reveal intrinsic clustering of GRBs by origin, morphology, and observational 
properties (SNR, duration). 
The geometric structure of the low-dimensional UMAP embedding 
reflects the model’s success in capturing fundamental physical differences between 
GRB subclasses.
Misclassified GRBs are predominantly located in low-SNR regions, suggesting 
that these faint detections require specialized further analysis. 
The model exhibits clear distribution differences between long and short bursts, 
confirming its ability to capture discriminative features for GRB classification 
and successfully capture subtle variations in light curve morphology and temporal 
evolution patterns.

We also gain valuable physical insights from model-derived features. 
The dimensionality reduction and AutoGMM clustering of model-extracted features 
uncover meaningful physical patterns in GRB populations. 
Kilonova-associated GRBs (GRB-KN) predominantly cluster in Clusters 2 and 3, 
with long-duration GRB-KN in Cluster 2 sharing multi-phase light curve 
morphologies—supporting a common progenitor system as proposed by previous 
theoretical work \citep{KN-GRB_Zhu2022ApJ}, suggesting a common progenitor system 
for these apparently long-duration events. 
A particularly compelling case is presented by GRB~211211A and GRB~230307A, 
which exhibit nearly identical three-phase emission structures 
(precursor, main burst, extended emission) and occupy adjacent regions in 
the low-dimensional embedding 
space \citep{DownDimGRB_Dimple2023, DownDimGRB_Negro2025}.
The comprehensive temporal and spectral analyses of their strikingly similar 
properties indicate a common 
mechanism \citep{GRB-KN-Compare_2024Peng, GRB-KN-Compare_2025Wang}. 
This clustering pattern mirrors yet remains distinct from the tight grouping 
of GRB-SN in Clusters 1 and 2, which similarly reflects their shared progenitor 
physics \citep{GRB-SN-similar_2024Kumar, GRB-SN-similar_2025Kumar}. 
GRBs with extended emission show spatial separation into three subgroups 
highly correlated with duration, while ultra-long GRBs are confined to the 
lGRB-dominated Cluster 3—hinting at potential subclasses with physical origins 
independent of standard duration-based classification.
Notably, sGRBs with precursors occupy transitional zones between short 
and long GRB clusters, indicating hybrid characteristics and diverse 
progenitor systems.
In contrast, lGRBs with precursors show widespread dispersion, implying 
precursor activity is an independent physical process rather than a core burst 
characteristic. 
These findings reinforce the limitations of duration-only GRB classification, 
aligning with prior studies (e.g., \citep{GRB-cluter_Modak_2021}) that highlight 
inherent uncertainties in this paradigm. 
The model’s ability to aggregate GRBs by physical origin rather than just 
observational parameters demonstrates that machine learning-derived features 
can serve as a complementary dimension for probing GRB progenitors, 
with future increases in the sample size of events with confirmed origins 
essential for further validating these classification schemes.

Despite achieving high classification accuracy, the model exhibits residual 
misclassification of low-SNR GRBs (99\% of false negatives have 
peak-SNR $\textless$ 5 $\sigma$), which may require specialized feature 
engineering or multi-modal data integration 
(e.g., combining light curves with spectral data, energy response-corrected data) 
for further improvement. 
Predictive uncertainty quantification—critical for real-world astrophysical 
applications—also warrants deeper integration, as confidence assessment for 
low-significance or atypical bursts remains underdeveloped. 
Additionally, validating the clustering results for special GRB subclasses 
(e.g., GRB-KN, precursor-containing GRBs) requires larger samples of events 
with confirmed origins.
Future work will focus on three key directions: 
(1) developing robust, real-time uncertainty estimation to deliver reliable confidence assessments for faint and atypical transients, which is critical for triggering follow-up observations. 
(2) expanding the framework to process multi-modal data (spectral, temporal, 
and polarization information) for richer feature extraction; 
(3) adapting the model for real-time analysis pipelines in time-domain and 
multi-messenger astronomy, enabling faster detection and characterization of 
transient phenomena. The adaptive frequency-domain approach and data 
augmentation strategy developed here are also generalizable to other 
transient astrophysical events (e.g., fast radio bursts, soft gamma-ray 
repeaters, supernovae), highlighting their broader potential for advancing 
observational capabilities.


In summary, this study advances GRB identification through a synergistic 
combination of physics-informed data augmentation and frequency-adaptive 
feature enhancement, effectively addressing core challenges of limited training 
samples, observational selection bias, and inadequate discriminative feature 
extraction in GRB studies. 
By generating over 100,000 physically faithful pseudo-GRB samples,  via our 
augmentation strategy, we effectively mitigate data scarcity, alleviate the
"observational iceberg" effect, and substantially enhance the model’s 
generalization capability to faint, low-significance bursts.
We further elevate detection performance by developing a ResNet-based deep 
learning framework integrated with the novel AFFE module, which adaptively 
weights and filters frequency-domain components to capture key GRB signatures. 
The framework not only achieves high classification accuracy that 
outperforming conventional attention mechanisms and frequency-domain approaches, 
but also provides actionable physical insights into GRB populations by 
capturing intrinsic relationships between observational signatures and
progenitor origins.
Through feature visualization and dimensionality reduction, we demonstrate 
that GRBs aggregated in the model’s feature space share similar physical 
attributes, highlighting that machine learning-derived features offer an 
additional dimension for probing GRB origins beyond traditional duration-based 
classification paradigms. 
We emphasize the broad potential of adaptive frequency-domain analysis for 
advancing GRB identification and underscore the generalizability of our 
approach to other transient astronomical phenomena. 
As time-domain and multi-messenger astronomy rapidly evolve, there is an 
urgent demand for higher accuracy, improved temporal resolution, 
and the ability to uncover latent feature relationships. 
Future work will integrate generalized machine learning models capable 
of processing multi-band and multi-modal data into real-time analysis 
pipelines—enabling deeper feature extraction, faster transient detection, 
and more precise characterization—thereby laying the groundwork for robust 
transient detection systems that leverage the growing volume of data from 
next-generation astronomical missions.

\section{Acknowledgements}
We would like to thank Prof Rui Luo, Dr. Yi Yang, Prof Jia-Wei Luo, and Rui-Ze Shi for 
helpful discussion. 
This study is supported by the National Key R\&D Program of China
(2024YFA1611703 and 2024YFA1611700), the National Natural Science Foundation of 
China (grant Nos. 12473044, 12494572, 12273042, 12133007, 41827807 and 61271351) 
and the Science and Technology Innovation Plan of Shanghai Science and Technology 
Commission (22DZ1209500). 
This work is also partially supported by the Strategic Priority Research Program 
of the CAS under grant
No. XDA15360300. B. L. acknowledges support from the National Astronomical
Science Data Center Young Data Scientist Program (grant No. NADC2023YDS-04).


The code and data sets are available upon reasonable request.

\bibliographystyle{aasjournal}
\bibliography{paper.bib}


\end{CJK*}
\end{document}